\documentclass[amssymb,superscriptaddress,prl,showpacs,twocolumn,tightenlines,balancelastpage,10pt,a4paper]{revtex4-1}
\usepackage{graphicx}
\usepackage{bm}
\usepackage{amsmath}
\usepackage[bookmarks=false]{hyperref}
\usepackage{dcolumn}
\usepackage{float}
\usepackage{ulem}

\usepackage[usenames,dvipsnames,svgnames,table]{xcolor}

\begin{document}

\title{Quantum gambling based on Nash-equilibrium}
\author{Pei Zhang}
\affiliation{MOE Key Laboratory for Nonequilibrium Synthesis and Modulation of Condensed Matter, Department of Applied Physics, Xi'an Jiaotong University, Xi'an 710049, China}
\affiliation{Centre for Quantum Photonics, H. H. Wills Physics
Laboratory \& Department of Electrical and Electronic Engineering,
University of Bristol, BS8 1UB, United Kingdom}
\author{Xiao-Qi Zhou}
\email{Xiaoqi.Zhou@bristol.ac.uk}
\affiliation{Centre for Quantum Photonics, H. H. Wills Physics
Laboratory \& Department of Electrical and Electronic Engineering,
University of Bristol, BS8 1UB, United Kingdom}
\author{Yun-Long Wang}
\affiliation{MOE Key Laboratory for Nonequilibrium Synthesis and Modulation of Condensed Matter, Department of Applied Physics, Xi'an Jiaotong University, Xi'an 710049, China}
\author{Peter J. Shadbolt}
\affiliation{Centre for Quantum Photonics, H. H. Wills Physics
Laboratory \& Department of Electrical and Electronic Engineering,
University of Bristol, BS8 1UB, United Kingdom}
\author{Yong-Sheng Zhang}
\affiliation{Key Laboratory of Quantum Information, University of Science and
Technology of China, CAS, Hefei 230026, China}
\author{Hong Gao}
\affiliation{MOE Key Laboratory for Nonequilibrium Synthesis and Modulation of Condensed Matter, Department of Applied Physics, Xi'an Jiaotong University, Xi'an 710049, China}
\author{Fu-Li Li}
\affiliation{MOE Key Laboratory for Nonequilibrium Synthesis and Modulation of Condensed Matter, Department of Applied Physics, Xi'an Jiaotong University, Xi'an 710049, China}
\author{Jeremy L. O'Brien}
\affiliation{Centre for Quantum Photonics, H. H. Wills Physics
Laboratory \& Department of Electrical and Electronic Engineering,
University of Bristol, BS8 1UB, United Kingdom}

\begin{abstract}
A fair gambling is hard to be made between two spatially separated parties without introducing a trusted third party. Here we propose a novel gambling protocol, which enables fair gambling between two distant parties without the help of a third party. By incorporating the key concepts and methods of game theory, our protocol will ¡°force¡± the two parties to move their strategies to a Nash-equilibrium point which guarantees the fairness through the physical laws of quantum mechanics. Furthermore, we show that our protocol can be easily adapted to a biased version, which would find applications in lottery, casino, etc. A proof-of-principle optical demonstration of this protocol is reported as well.

\end{abstract}


\maketitle


Gambling---a game in which people wager money or something valuable on an event with an uncertain outcome---has a wide range of applications in every aspects of human society \cite{walker92}. However, despite its long history and wide spread usages, it has a long standing problem yet to be resolved. Suppose a gambler~(say Bob) wants to gamble with the Casino~(say Alice), how does Bob know the gambling machine (GM) provided by Alice is not biased towards Alice herself, especially in the case of online gambling or lotteries?

The standard solution to this problem is to introduce a trusted third party to provide an unbiased GM to make sure the gambling is fair to both parties. However, in some cases such third party which is trusted by both parties does not exist. Surprisingly, by drawing from the classical \cite{nash} and quantum game theory \cite{meyer99,Eisert99}, we have found a protocol which enables two parties to create an unbiased GM themselves to perform a fair gambling without introducing any third party. The GM, which has two independent parameters,  is constructed by Alice and Bob together who  can change the values of the two parameters respectively.
Furthermore, the GM is elaborately designed in a way that a Nash-equilibrium~\cite{nash} exists---each party has a strategy to choose his/her parameter which can guarantee his/her gain is no less than a certain amount and neither of the two parties can benefit from changing his/her own
parameter unilaterally. In this way, Alice and Bob are `forced' to choose the Nash-equilibrium
in their own favor so that a stable GM can be established.

The paper is structured as follows. First, we will describe the protocol in detail and explain how the Nash-equilibrium can guarantee the GM to be unbiased to each party. Then, we will show how to generalize this protocol to a full family of quantum gambling, including both biased and unbiased,
by introducing several parameters.
At last, we present a proof-of-principle optical experiment to demonstrate the protocol.

The rules of the game and the strategies the players should follow are explained below.

\vspace{1mm}
\noindent \textbf{The rules of the game:} Alice has two boxes, named $A$ and $B$, which are used to store a particle. The quantum states of the particle stored in the two boxes are denoted $|a\rangle$ and $|b\rangle$, respectively. Alice prepares the particle in a state and then sends the box $B$ to Bob. Bob wins in one of the following cases: (1) Bob opens the box $B$ and find the particle. (2) Bob does not find the particle in box $B$ and asks Alice to send him the box $A$. Bob then detects the state Alice prepared is different from the committed state $|\psi_{c}\rangle$, where $|\psi_{c}\rangle=\ \frac{1}{3} |a\rangle \ +\ \frac{2\sqrt{2}}{3} |b\rangle$. In any other cases, Alice wins.

\vspace{3mm}
 \noindent \textbf{Alice's strategy:} Alice prepares
the particle in the following state:
\begin{equation}
| \psi \rangle  \ = \ \sqrt{1- \alpha} |a\rangle \ +\ \sqrt{\alpha} |b\rangle,
\end{equation}
where $\alpha$ is a parameter controlled by Alice ($0\leqslant\alpha\leqslant1$).

\vspace{3mm}
\noindent \textbf{Bob's strategy:}

After receiving the box $B$, Bob splits the particle into two parts. One part is still stored in box $B$ and the other part is stored in a new box $B'$. Specifically, Bob performs the following operation:
Bob splits the box into two parts, $B$ and $B'$:
\begin{equation}
|b\rangle \longrightarrow  \sqrt{1-\beta}|b\rangle+\sqrt{\beta}|b'\rangle,
\end{equation}
where $|b'\rangle$ denotes the quantum state of the particle stored in the box $B'$. The splitting ratio $\beta$ is a parameter controlled by Bob ($0 \leqslant \beta \leqslant 1$).
After the splitting, Bob opens the box $B$ and measures the projection operator on the state $|b\rangle$. If he finds the particle in the box $B$, he wins.
If he doesn't, he asks Alice for the box $A$ and combines it with the box $B'$ to make a verification. If the verification shows the initial state Alice prepared is different from the committed state $|\psi_{c}\rangle$, Bob still wins; otherwise, Alice wins.

This completes the definition of our protocol.

\begin{figure*}[t]
\includegraphics[width=0.975\textwidth]{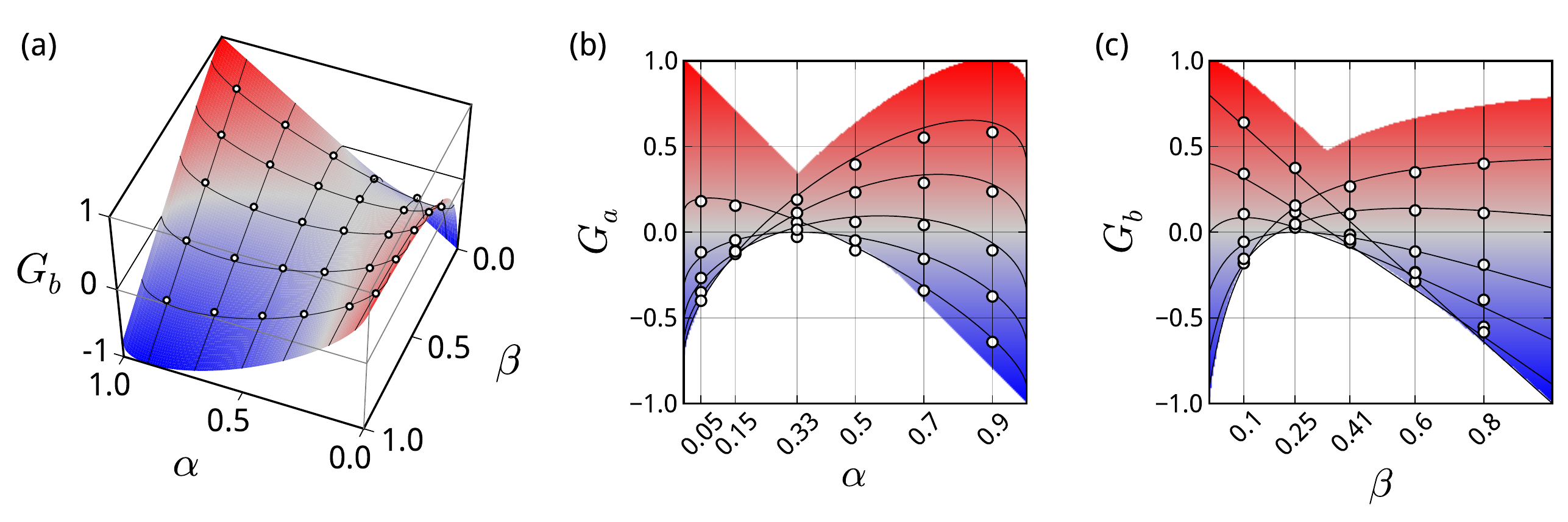}
\caption{Theoretical (surface) and experimental (circles) results of our protocol under $R=1$ and $\gamma=8/9$. (a) Bob's average gain in a three dimensional view. The Nash-equilibrium is the point of $G_{b}=0$, $\beta=1/4$ and $\alpha=1/3$. (b) Bob's gain under his parameter $\beta$. The best strategy is choosing $\beta=1/4$.  (c) Alice's gain under her parameter $\alpha$. Her best choice is $\alpha=1/3$. }
\end{figure*}

Let us briefly analyze the protocol and both players' strategies. For Alice, she has a motivation to prepare a state that the particle has a higher chance to stay in box $A$ (choosing a small $\alpha$), so that Bob has a lower chance to find the particle in box $B$. However, if $\alpha$ is too small, the discrepancy between the prepared and the committed state would be too big which will result in a higher chance for Alice to lose in the verification stage---there is a tradeoff for Alice to choose her strategy~(parameter $\alpha$). Similar analysis can apply to Bob's strategy as well. Bob can't rely solely on the $|b\rangle$ projection stage~(choosing a small $\beta$) or the verification stage~(choosing a big $\beta$). He needs to consider both stages and choose an intermediate $\beta$ to maximize his chance to win. By now, the analysis is just qualitative. Actually we have found that there exists best strategies for both parties---the best strategy for Alice~(Bob) is to set $\alpha=\frac{1}{3}$~($\beta = \frac{1}{4}$). The average gain of Alice~(Bob) $G_{a}$~($G_{b}$) will never be negative once she~(he) chooses $\alpha$~($\beta$) to be $\frac{1}{3}$~($\frac{1}{4}$), where the equation $G_{a}+G_{b}=0$ holds as the gambling is a zero-sum game. When both of them choose their best strategies, $G_{a}$ and $G_{b}$ have to be zero and a fair gambling will be achieved.

To prove the above claims and features of the protocol, let us write down the expression for $G_{b}$ first \cite{note},
\begin{equation}
G_{b}=P_{1}+P_{2}-(1-P_{1}-P_{2}),
\end{equation}
where $P_{1}$ and $P_{2}$ denote the probability for Bob to find the particle in box $B$ and to find the initial state is different with committed state, respectively.
When Bob receives the box $B$ and splits one part to the box $B'$, the state becomes
\begin{equation}
| \psi_0 \rangle  \ = \ \sqrt{1- \alpha} |a\rangle \ + \sqrt{\alpha} (\sqrt{1-\beta} |b\rangle + \sqrt{\beta} |b'\rangle).
\label{psi0}
\end{equation}
From Eq. \ref{psi0}, it is straightforward to calculate the probability ($P_{1}$) of finding the particle in box $B$,
\begin{equation}
P_1=\| \langle\ b| \psi_0 \rangle \|^2 =\alpha  (1-\beta).
\end{equation}
The state of the particle will collapse to $| \psi_0^\prime \rangle$ if Bob fails to detect the particle in box $B$, where
\begin{equation}
| \psi_0^\prime \rangle  \ =\sqrt{ \frac{1-\alpha}{1-\alpha+\beta\alpha}}|a\rangle  + \sqrt{\frac{\beta\alpha}{1-\alpha+\beta\alpha}}|b'\rangle,
\end{equation}
If Alice did prepare the particle in the committed state $| \psi_c\rangle$ initially, the state at this stage will be $| \psi_c ^\prime \rangle$
\begin{equation}
| \psi_c ^\prime \rangle  \ = \sqrt{ \frac{1}{1+8\beta}}|a\rangle  + \sqrt{\frac{8\beta}{1+8\beta}}|b'\rangle.
\end{equation}
Bob then makes a projection measurement on $|\psi_{c}^{\prime}\rangle$ for the verification. If the outcome is negative, Bob knows with certainty that state Alice prepared is different with the committed state  $|\psi_{c}\rangle$. The probability of detecting such event is given by
\begin{eqnarray}
P_{2} &=& (1-P_1)(1-\|\langle \psi_{c}^{\prime} |\psi_{o}^{\prime}\rangle \|^{2})\notag\\
 &=& \frac{\beta[8-7\alpha-4\sqrt{ 2 \alpha(1-\alpha)}]}{1+8\beta}
\end{eqnarray}
By substituting Eqs. (5) and (8) into Eq. (3), we can get $G_{b}$ as a function of $\alpha$ and $\beta$:
\begin{eqnarray}
&&G_{b}(\alpha,\beta)=\frac{1}{1+8\beta} \{2\alpha+8\beta-1\notag\\
&&\ \ \ \ \ \ \ \ \ \ \ -8[2\beta^2\alpha-\beta\sqrt{2 \alpha(1-\alpha)} ]\}.
\end{eqnarray}

As shown in Fig. 1a, the $G_{b}$ function is saddle-shaped and the saddle point is at $\alpha=\frac{1}{3}$ and $\beta=\frac{1}{4}$. Figures 1b and 1c are the projection of the $G_{b}$ function to the $\beta-G_{b}$ plane and $\alpha-G_{a}$ plane, respectively. From Fig. 1b~(1c), it's clear that, no matter what strategy Alice~(Bob) chooses, Bob's~(Alice's) gain will always be non-negative if he~(she) sets his~(her) parameter to be $\frac{1}{4}$~($\frac{1}{3}$). Any party changes his/her strategy unilaterally will only decrease his/her own gain. In this way, Nash-equilibrium is achieved and both parties will stick their strategy and thus a stable and fair game is achieved.

The above scheme effectively realizes a coin-tossing protocol \cite{qcf1,qcf2,qcf3,qcf4,qcf5,qcf6,qcf7,qcf8} where the one-shot gains for both parties are balanced and the expectation value of the gains for both parties are zero. However, in practice, there are much more diverse gambling protocols, such as Roulette or Lottery, where the one-shot gains are unbalanced and the expectation value of the gains are non-zero. As only the ratio of the one-shot gains matters, without loss of generality, we fix Alice's one-shot gain to be 1 and Bob's one-shot gain to be $R$.
Now the task is to design a stable gambling protocol in which Bob's one-shot gain is $R$ and the expectation value of Bob's gain $G_{b}$ to be $\delta$. It can be proved that our scheme can be easily extended to realize such gambling protocol by just reappointing the committed state to be
\begin{equation}
|\psi_{c}\rangle=\ \sqrt{1- \gamma} |a\rangle \ +\ \sqrt{\gamma} |b\rangle,
\label{eq12}
\end{equation}
where
\begin{equation}
\gamma=\frac{4(1+\delta)(1+R)}{(2+\delta+R)^2}.
\label{eq14}
\end{equation}
The Nash-equilibrium point for Alice's and Bob's strategies would be $\alpha=\frac{1-\sqrt{1-\gamma}}{2}$ and $\beta=\frac{-1+\gamma+\sqrt{1-\gamma}}{\gamma}$. Following similar reasoning shown in last paragraph, setting $\alpha$ and $\beta$ to the above values are the best strategies for Alice and Bob, as none of them can benefit from changing their own strategy unilaterally. The earlier protocol can be regarded as a special case where $R=1$, $\delta=0$ and $\gamma=8/9$. The detailed calculation and proof can be found in the Appendix.

Besides proposing the theory, we also implemented a proof-of-principle experiment to demonstrate our gambling protocols. As shown in Fig. 2,
a He-Ne laser centred at 632.8 nm is attenuated to the level of $\sim$100 KHz count rate to serve as single-photon source.
The two polarization states $| V \rangle$ and $| H \rangle$ are encoded as the two box states $| a \rangle$ and $| b \rangle$, respectively. Thus the two parameters $\alpha$ and $\beta$ can be easily managed by half-wave-plates (HWP). HWP$_1$ is controlled by Alice to prepare
the state $| \psi \rangle$. Bob uses HWP$_2$ and polarized beam splitter (PBS$_2$) to split the state $| b \rangle$ to $| b \rangle$ and $| b' \rangle$, and measures $P_1$ at the single-photon detector (D$_1$). Then $| b' \rangle$ and $| a \rangle$ are combined at PBS$_1$ for verification.

A Sagnac interferometer~(with interference visibility of 96\%) is used in our setup to make sure the phase stability is maintained throughout the whole experiment. HWP$_3 $ and PBS$_3$ are used for the projective measurement, where $P_2$ and $P_3$ can be measured from D$_2$ and D$_3$, respectively.

In our experiment, we simulated a fair coin-tossing GM, where $R$ and $\gamma$ are set to be $1$ and $8/9$, respectively. Both Alice and Bob chose a series of strategies and the final gains for both parties are measured and recorded. The results are shown in Fig. 1. All the data are well agreed with the theoretical predictions. From the results, we can clearly see that the best gain Alice~(Bob) can get is when she~(he) choose the strategy $\alpha=1/3$~($\beta=1/4$). For their own good, Alice and Bob would both choose their best strategies and thus a Nash-equilibrium is formed and a fair gamble is achieved.

\begin{figure}[h]
\includegraphics[width=0.4\textwidth]{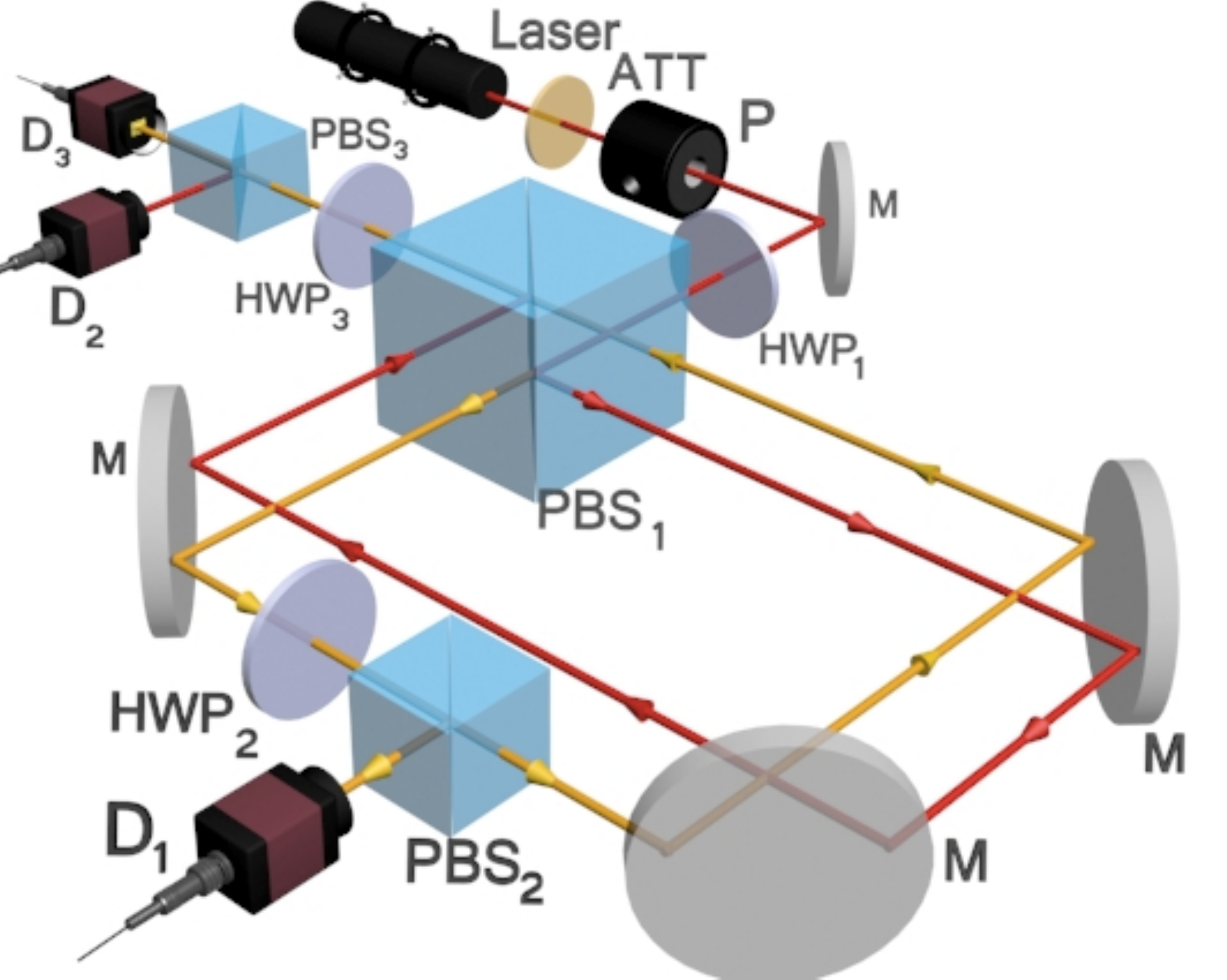}
\caption{Experimental demonstration of quantum gambling. ATT represents attenuator which deeply attenuates the laser to the single photon level. P is a polarizer. PBS is the acronym for polarized beam splitter which transmitting horizontal polarized light and reflecting vertical polarized light. HWP indicates the half-wave-plate. HWP$_1$ is controlled by Alice as the parameter $\alpha$; HWP$_2$ and PBS$_2$ are inserted in the counter-clockwise route (green line) which are controlled by Bob as the parameter $\beta$; HWP$_3 $ is chosen for projective measurement and acts as the parameter $\gamma$. We use three single-photon detectors D$_1$, D$_2$, and D$_3$ to get the three probabilities $P_1$, $P_2$, and $P_3$, respectively. The polarization-Sagnac interferometer ensures a stable system to collect the data.}
\end{figure}

The errors of our experimental results mainly come from the imperfection of components used in the setup. In our protocol of GM, although $R$ can be chosen freely, we recommend to set a small value of $R$ as the error of the gain $G_{b}$ is proportional to $R$ (see Eq.~(\ref{Gb0}) in the Appendix).

Let us now consider the possible ``cheating'' strategies Alice and Bob might use in this protocol. For Alice, if she prepares the particle not only in boxes $A$ and $B$ but also in another box $C$, this will only increase the probability of Bob finding the state different from committed state, which has no benefit to Alice herself. Introducing ancillary particles does not help Alice either. So, for Alice, there is no meaningful cheating strategy. However, for Bob, he might try to cheat. He can claim he find the initial state different with the committed state even if the verification result shows the opposite. Alice cannot tell if Bob is lying or not when the initial state is not the same as the committed state. To detect this cheating, Alice can occasionally prepare the particle in the commited state. However, this detecting procedure will decrease Alice's average gain. To compensate this loss to Alice, the $G_{a}$ on the Nash-equilibrium point can be set to a value slightly greater than zero. This can be easily achieved by changing $\gamma$ or $R$ as shown in our protocol.

We note that there are several other quantum protocols~\cite{Gold1999, zhang08, hwang01, hwang02} designed for gambling without third party. However, none of them can achieve a fair game between the casino and the gamblers.


In summary, we have invented a protocol which can promise an unbiased GM to each party by using quantum gambling theory and Nash-equilibrium. Furthermore, the choice of parameter values is flexible, and we have found the relationship between these adjustable parameters, which can be used to guide a feasible implementation of full family of quantum gambling, including both biased and unbiased cases. Compare with related protocols such as quantum coin flipping \cite{qcf1,qcf2,qcf3,qcf4,qcf5,qcf6,qcf7,qcf8} and quantum bit commitment \cite{bc1,bc2,bc3,bc4,bc5,bc6,bc7,bc8,bc9}, the key difference is that quantum gambling makes the cheating as part of strategy, and set flexible rewards to affect the choice of casino and player. This proof-of-principle experiment therefore provides solid support for the applicability and feasibility of our scheme. In a world full of competitions and cooperations, we believe our protocol of gambling machine without a third party will provide direct applications in the near future, and also shed light on developing new quantum technologies.

This work is supported by the Fundamental Research Funds for the Central Universities,
National Natural Science Foundation of China (Grant Nos. 11004158, 11374008, 11074198, and 60778021), EPSRC, ERC, QUANTIP, PHORBITECH, and NSQI. J. L. OB. acknowledges a Royal Society Wolfson Merit Award and a Royal Academy of Engineering Chair in Emerging Technologies.

\section{Appendix}
The rules of the protocol can be modified to be general.

\vspace{3mm}
\noindent \textbf{The rules of the game:}
Alice can store a quantum particle in two boxes, $A$ and $B$, and the state of the particle is denoted by $|a\rangle$ and $|b\rangle$, respectively. However, she only sends one box (suppose box $B$) to Bob, who will open the box and check where the particle is. Alice and Bob consent to use a superposition state of $|a\rangle$ and $|b\rangle$ as the conventional state. We can define the conventional state to be
\begin{equation}
|\psi_{c}\rangle=\ \sqrt{1- \gamma} |a\rangle \ +\ \sqrt{\gamma} |b\rangle,
\end{equation}
Bob wins $R$ coins in one of two cases: finding the particle in box $B$, or detecting that the state prepared by Alice is different from the conventional state $|\psi_{c}\rangle$. Otherwise, Otherwise, he loses 1 coin to Alice.

The strategies for them are same as we described in the main text.
So the whole game can be acted as this: Alice and Bob choose a state as conventional state $|\psi_{c}\rangle$ before start. Then Alice chooses a parameter $\alpha$ to prepare a state of Eq. (2), and sends box $B$ to Bob. In Bob's side, he chooses a splitting parameter $\beta$ and splits the box $B$ into boxes $B$ and $B'$. If Bob gets the particle in $B$, he wins $R$ coins (after Alice check the box $A$); If $B$ is empty, he asks Alice to send him box $A$ and combines it with box $B'$ to make a verification, then he may win $R$ coins or lose $1$ coin depending on the verification results show $|\psi\rangle\neq|\psi_{c}\rangle$ or $|\psi\rangle=|\psi_{c}\rangle$, respectively.

Following the rules and the strategies of the game, the expectation value of Bob's gain is
\begin{equation}
G_{b}=R(P_{1}+P_{2})-P_{3}.
\label{Gb0}
\end{equation}
The state before Bob makes a detection can be written as
\begin{equation}
| \psi_0 \rangle  \ = \ \sqrt{1- \alpha} |a\rangle \ + \sqrt{\alpha} (\sqrt{1-\beta} |b\rangle + \sqrt{\beta} |b'\rangle).
\end{equation}
$P_{1}$ is the probability of finding particle in box $B$ and it in turn leads to
\begin{equation}
P_1=\| \langle\ b| \psi_0 \rangle \|^2 =\alpha  (1-\beta).
\end{equation}
The state for verification is
\begin{equation}
| \psi_0^\prime \rangle  \ =\sqrt{ \frac{1-\alpha}{1-\alpha+\beta\alpha}}|a\rangle  + \sqrt{\frac{\beta\alpha}{1-\alpha+\beta\alpha}}|b'\rangle,
\end{equation}
while the expecting state to be checked should be
\begin{equation}
| \psi_c ^\prime \rangle  \ = \sqrt{ \frac{1-\gamma}{1-\gamma+\beta\gamma}}|a\rangle  + \sqrt{\frac{\beta\gamma}{1-\gamma+\beta\gamma}}|b'\rangle.
\end{equation}
If Bob does not find the particle in box $B$, he will project $|\psi_{o}'\rangle$ onto $|\psi_{c}^{\prime}\rangle$. Then we can get
\begin{eqnarray}
&&P_{3}=(1-P_1)\|\langle \psi_{c}^{\prime} |\psi_{o}^{\prime}\rangle \|^{2} \notag \\
&&\ \ \ \ \ = \frac{(\sqrt{(1-\alpha)(1-\gamma)}+\beta\sqrt{\gamma\alpha})^2}{1-\gamma+\beta\gamma}, \\
&&P_2=1-P_1-P_3 \notag \\
&&\ \ \ \ \ =\frac{\beta[\gamma+\alpha-2\gamma\alpha-2\sqrt{ \gamma \alpha(1-\alpha)(1-\gamma)}]}{1-\gamma+\beta\gamma}  .
\end{eqnarray}
Substituting Eqs. (6), (9) and (10) into Eq. (4), we can get $G_{b}$ is in the form of four parameters: $\alpha$, $\beta$, $\gamma$ and $R$:
\begin{eqnarray}
&&G_{b}=\frac{1}{1-\gamma+\beta\gamma} \{R(\alpha-\gamma \alpha+\beta\gamma)-(1-\alpha)(1-\gamma)\notag\\
&&\ \ \ \ \ \ \ \ \ \ \ -(1+R)[\beta^2\gamma\alpha-2 \beta\sqrt{\gamma \alpha(1-\alpha)(1-\gamma)} ]\}.
\label{Gb}
\end{eqnarray}



To proof there is a Nash-equilibrium point in the Eq.~(\ref{Gb}), we use the definition of Nash-equilibrium. In order to find the best strategy for Bob, we should first minimize $G_{b}$ for $\alpha$ and then maximize the result for $\beta$. This means that no matter what strategy Alice chooses, Bob can make sure his gain is no less than a value $\delta$. To find the best strategy for Alice, we should first maximize $G_{b}$ for $\beta$ and then minimize the result for $\alpha$. The calculation yields that the Nash-equibrilium is at
\begin{eqnarray}
\label{eq14}
&&\delta=\frac{2+2 R-\gamma (2+R)-2 (1+R) \sqrt{(1-\gamma)}}{\gamma};\\
\label{eq27}
&& \alpha=\frac{1-\sqrt{1-\gamma}}{2};  \\
\label{eq28}
&& \beta=\frac{-1+\gamma+\sqrt{1-\gamma}}{\gamma}.
\end{eqnarray}


From Eq.~(\ref{eq14}), we can know how to choose the parameter of conventional state:
\begin{equation}
\label{eq29}
\gamma=\frac{4(1+\delta)(1+R)}{(2+\delta+R)^2}.
\end{equation}
Equations~(\ref{eq14}), (\ref{eq27}), (\ref{eq28}), and (\ref{eq29}) are the results for general quantum gambling.


\begin{thebibliography}{99}

\bibitem{walker92} M. B. Walker, \textit{The psychology of gambling} (Pergamon Press, Oxford and New York, 1992).

\bibitem{nash} J. F. Nash, Proc. Natl. Acad. Sci. USA \textbf{36}, 48-49 (1950).


\bibitem{meyer99} D. Meyer, Phys. Rev. Lett. \textbf{82}, 1052 (1999).

\bibitem{Eisert99} J. Eisert, M. Wilkens, and M. Lewenstein, Phys. Rev. Lett. \textbf{83}, 3077 (1999).

\bibitem{note} For it is a zero-sum game, $G_{a}=-G_{b}$, without loss of generality, we can only calculate Bob's average gain $G_{b}$.

\bibitem{qcf1} R. W. Spekkens and T. Rudolph, Phys. Rev. A \textbf{65}, 012310 (2001).

\bibitem{qcf2} R. W. Spekkens and T. Rudolph, Phys. Rev. Lett. \textbf{89}, 227901 (2002).

\bibitem{qcf3} C. D\"oscher, and M. Keyl, Fluct. Noise Lett. \textbf{2}, R125 (2002).

\bibitem{qcf4} C. Mochon, Phys. Rev. A \textbf{72}, 022341 (2005).

\bibitem{qcf5} N. Aharon and J. Silman, New J. Phys. \textbf{12}, 033027 (2010).

\bibitem{qcf6} G. Berl\'{\i}n, G. Brassard, F. Bussi\'{e}res, and N. Godbout, Phys. Rev. A \textbf{80}, 062321 (2009).

\bibitem{qcf7} G. Berl\'{\i}n, G. Brassard, F. Bussi\'{e}res, N. Godbout, J. A. Slater, and W. Tittel, Nat. Commun. \textbf{2}, 561 (2011).

\bibitem{qcf8} J. Silman, A. Chailloux, N. Aharon, I. Kerenidis, S. Pironio, and S. Massar, Phys. Rev. Lett. \textbf{106}, 220501 (2011).

\bibitem{Gold1999} L. Goldenberg, L. Vaidman, and S. Wiesner, Phys. Rev. Lett. \textbf{82}, 3356 (1999).

\bibitem{zhang08} P. Zhang, Y.-S. Zhang, Y.-F. Huang, L. Peng, C.-F. Li, and G.-C. Guo, Europhys. Lett. \textbf{82}, 30002 (2008).







\bibitem{hwang01} W. Y. Hwang, D. Ahn, and S. W. Hwang, Phys. Rev. A \textbf{64}, 064302 (2001).

\bibitem{hwang02} W. Y. Hwang and K. Matsumoto, Phys. Rev. A \textbf{66}, 052311 (2002).

\bibitem{bc1} D. Mayers, Phys. Rev. Lett. \textbf{78}, 3414 (1997).
\bibitem{bc2} H.-K. Lo and H. F. Chau, Phys. Rev. Lett. \textbf{78}, 3410 (1997).
\bibitem{bc3} A. Kitaev, D. Mayers, and J. Preskill, Phys. Rev. A \textbf{69}, 052326 (2004).
\bibitem{bc4} G. M. DÕAriano, D. Kretschmann, D. Schlingemann, and R. F. Werner, Phys. Rev. A \textbf{76}, 032328 (2007).

\bibitem{bc5} A. Kent, New J. Phys. \textbf{13}, 113015 (2011).
\bibitem{bc6} A. Kent, Phys. Rev. Lett. \textbf{109}, 130501 (2012).
\bibitem{bc7} N. H. Y. Ng, S. K. Joshi, C. C. Ming, C. Kurtsiefer, and S. Wehner, Nat. Commun. \textbf{3}, 1326 (2012).
\bibitem{bc8} J. Kaniewski, M. Tomamichel, E. HŠnggi, and S. Wehner, IEEE Trans. Inf. Theory \textbf{59}, 4687 (2013).
\bibitem{bc9} Y. Liu, Y. Cao, M. Curty, S.-K. Liao, J. Wang, K. Cui, Y.-H. Li, Z.-H. Lin, Q.-C. Sun, D.-D. Li, H.-F. Zhang, Y. Zhao, T.-Y. Chen, C.-Z. Peng, Q. Zhang, A. Cabello, and J.-W. Pan, Phys. Rev. Lett. \textbf{112}, 010504 (2014).

\end{thebibliography}
\end{document}